\documentclass[conference]{IEEEtran}
\IEEEoverridecommandlockouts
\usepackage{cite}
\usepackage{amsmath,amssymb,amsfonts}
\usepackage{algorithmic}
\usepackage{graphicx}
\usepackage{textcomp}
\usepackage{xcolor}
\usepackage{booktabs}
\usepackage[hidelinks]{hyperref}

\newcommand{\upcite}[1]{\textsuperscript{\cite{#1}}}

\renewcommand{\figurename}{Figure}
\renewcommand{\tablename}{Table}

\def\BibTeX{{\rm B\kern-.05em{\sc i\kern-.025em b}\kern-.08em
    T\kern-.1667em\lower.7ex\hbox{E}\kern-.125emX}}
\begin{document}

\title{Tiny-Critic RAG: Empowering Agentic Fallback with Parameter-Efficient Small Language Models}

\author{\IEEEauthorblockN{1\textsuperscript{st} Yichao Wu*}
\IEEEauthorblockA{\textit{Northeastern University} \\
\textit
Boston, USA, 02115 \\
wu.yicha@northeastern.edu}
\and
\IEEEauthorblockN{2\textsuperscript{nd} Penghao Liang}
\IEEEauthorblockA{\textit{Northeastern University} \\
\textit
Boston, USA, 02115 \\
liang.p@northeastern.edu}
\and
\IEEEauthorblockN{3\textsuperscript{rd} Yafei Xiang}
\IEEEauthorblockA{\textit{Northeastern University} \\
\textit
Boston, USA, 02115 \\
xiang.yaf@northeastern.edu}
\and
\IEEEauthorblockN{4\textsuperscript{th} Mengwei Yuan}
\IEEEauthorblockA{\textit{Independent Researcher} \\
\textit
Milpitas, USA, 95035 \\
yuanmw1998@gmail.com}
\and
\IEEEauthorblockN{5\textsuperscript{th} Jianan Liu}
\IEEEauthorblockA{\textit{Independent Researcher} \\
\textit
Austin, USA, 78613 \\
jiananliu2408@gmail.com}
\and
\IEEEauthorblockN{6\textsuperscript{th} Jing Yang}
\IEEEauthorblockA{\textit{Washington University in St. Louis} \\
\textit
St. Louis, USA, 63130 \\
jing.y@wustl.edu}
\and
\IEEEauthorblockN{7\textsuperscript{th} Xianyou Li}
\IEEEauthorblockA{\textit{New York University} \\
\textit
New York, USA, 10012\\
xl4230@nyu.edu}
\and
\IEEEauthorblockN{8\textsuperscript{th} Weiran Yan}
\IEEEauthorblockA{\textit{Independent Researcher} \\
\textit
Milpitas, USA, 95035 \\
yanwr2016@gmail.com}
}

\maketitle

\begin{abstract}
Retrieval-Augmented Generation (RAG) grounds Large Language Models (LLMs) to mitigate factual hallucinations. Recent paradigms shift from static pipelines to Modular and Agentic RAG frameworks, granting models autonomy for multi-hop reasoning or self-correction. However, current reflective RAG heavily relies on massive LLMs as universal evaluators. In high-throughput systems, executing complete forward passes for billion-parameter models merely for binary routing introduces severe computational redundancy. Furthermore, in autonomous agent scenarios, inaccurate retrieval causes models to expend excessive tokens on spurious reasoning and redundant tool calls, inflating Time-to-First-Token (TTFT) and costs. We propose Tiny-Critic RAG, decoupling evaluation by deploying a parameter-efficient Small Language Model (SLM) via Low-Rank Adaptation (LoRA). Acting as a deterministic gatekeeper, Tiny-Critic employs constrained decoding and non-thinking inference modes for ultra-low latency binary routing. Evaluations on noise-injected datasets demonstrate Tiny-Critic RAG achieves routing accuracy comparable to GPT-4o-mini while reducing latency by an order of magnitude, establishing a highly cost-effective paradigm for agent deployment.
\end{abstract}


\section{Introduction}

\begin{figure*}[htbp]
\centerline{\includegraphics[width=\textwidth]{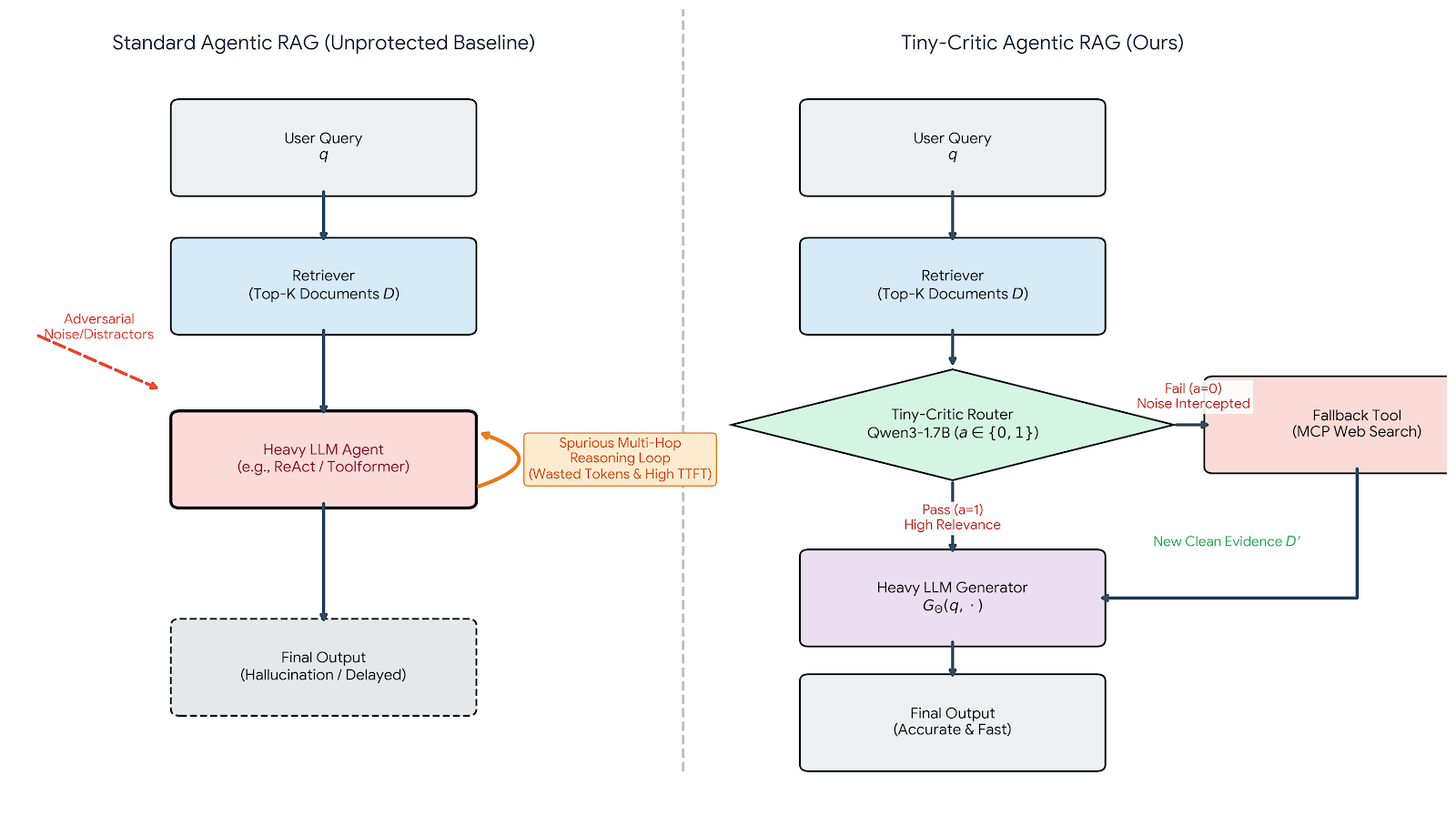}}
\caption{Architectural comparison highlighting the implicit cost of noise. \textbf{Left:} In an unprotected Agentic RAG (e.g., ReAct), adversarial noise triggers spurious multi-hop reasoning spirals, causing catastrophic token waste and high TTFT. \textbf{Right:} Our Tiny-Critic Agentic RAG preemptively intercepts noise ($a=0$), routing the query to a fallback tool to retrieve clean evidence ($D'$), effectively isolating the generator from hallucinations.}
\label{fig:architecture}
\end{figure*}

Retrieval-Augmented Generation (RAG) mitigates factual hallucinations in LLMs by grounding generation in verifiable corpora \upcite{b1, b2}. While naive RAG relies on static pipelines, recent paradigms shift towards Agentic architectures. These architectures introduce self-reflection mechanisms to handle noisy retrievals, enabling models to critique context before generation \upcite{b3, b4}. 

Crucially, in autonomous agent scenarios (e.g., ReAct or Toolformer), feeding inaccurate evidence triggers a cascading failure. As illustrated in Figure \ref{fig:architecture} (left), agents attempt to reconcile erroneous information, expending computational tokens on spurious reasoning steps and executing redundant tool calls \upcite{b6, b7}. This implicit multi-hop hallucination dilutes self-attention focus, drastically inflating TTFT and operational costs. Thus, an efficient preemptive evaluation mechanism is a structural imperative.

Despite improving system robustness, existing reflective frameworks predominantly deploy massive LLMs (e.g., GPT-4) as universal evaluators \upcite{b8}. In industrial retrieval systems with high concurrency, this heavyweight approach creates severe bottlenecks. Conversely, Superfiltering research demonstrates that SLMs can effectively act as gatekeepers for significantly larger models \upcite{b9}.

To resolve this latency-accuracy tension, we propose Tiny-Critic RAG (Figure \ref{fig:architecture}, right). Our framework decouples evaluation by introducing a parameter-efficient SLM (Qwen-1.7B) fine-tuned via LoRA as a dynamic gatekeeper \upcite{b10, b11}. Implementing hardware-aware constrained decoding \upcite{b12}, Tiny-Critic maps continuous evaluation into discrete routing actions, triggering fallbacks via Model Context Protocols (MCP) \upcite{b13} only when evidence is deficient. This curtails both explicit evaluation costs and implicit reasoning spirals.

\section{Related Work}

\subsection{Evolution of Agentic RAG and Reflection}
Naive RAG established baseline retrieval but suffered from distractors \upcite{b1}. Self-RAG enabled post-hoc reflection but required heavy generators to process entire sequences, wasting FLOPs \upcite{b3}. To preemptively filter noise, Corrective RAG (CRAG) introduced explicit retrieval evaluators \upcite{b4}. However, CRAG relied on resource-intensive models (e.g., T5-Large) for classification. Adaptive-RAG introduced complexity-based branching \upcite{b5}. Despite these advancements, existing frameworks fail to optimize the evaluator to the extreme constraints of a localized SLM with constrained decoding.

\subsection{Parameter-Efficient Tuning and Lightweight Critics}
Deploying specialized SLMs is supported by advances in parameter-efficient fine-tuning (PEFT) \upcite{b10, b15}. I-LoRA demonstrates routing-tuned adapters managing multi-task selections dynamically \upcite{b14}. Furthermore, process reward models (PRM) utilize lightweight critics to score intermediate trajectories, proving verification is computationally cheaper than generation \upcite{b20}. Integrating tool-use capabilities into SLMs enables autonomous execution without massive bases \upcite{b7}.

\section{Methodology}

\subsection{Problem Formulation and DAG Routing State Space}
Let $q$ denote a query and $\mathcal{C}$ the corpus. A retriever extracts evidence $D = \{d_1, \dots, d_k\} \subset \mathcal{C}$ \upcite{b18}. Standard RAG paradigms compute response probability $P(y \mid q, D)$ directly via generator $G_{\Theta}$.

In the Tiny-Critic DAG, we define a parameterized routing function $\pi_\phi(a \mid q, D)$ governed by an SLM. Similar to resilient routing frameworks in complex physical networks that utilize risk-aware mechanisms to bypass congestion \upcite{b21}, Tiny-Critic acts as a semantic state truncator with a binary action space $\mathcal{A} = \{0, 1\}$:
\begin{itemize}
\item \textbf{Generation Path ($a=1$):} If $D$ exhibits high semantic relevance, the system proceeds to compute $y = G_{\Theta}(q, D)$.
\item \textbf{Fallback Path ($a=0$):} If $D$ contains contradictory distractors, the system intercepts the workflow, executing tool $T_{fallback}(q)$ via MCP to yield clean context $D'$. Generation proceeds as $y = G_{\Theta}(q, D \cup D')$.
\end{itemize}

\subsection{Low-Rank Adaptation for Boundary Formulation}
To configure the SLM for routing without catastrophic forgetting, we apply LoRA \upcite{b10, b15}. For pre-trained matrix $W_0 \in \mathbb{R}^{d \times m}$, the update is $W = W_0 + BA$, with intrinsic rank $r \ll \min(d, m)$. Formatting input $x = \text{Concat}(q, D)$ and labels $y_{t} \in \{t_{pass}, t_{fail}\}$, the model optimizes cross-entropy loss over the final token:
\begin{equation}
\mathcal{L}(\phi) = -\sum_{i=1}^{N} \log P(y_{t}^{(i)} \mid x^{(i)}; W_0 + BA) \label{eq:loss}
\end{equation}

\subsection{Inference Acceleration via Constrained Decoding}
To strictly bound latency $L_{critic}$, we bypass autoregressive sampling. Utilizing the SLM's Non-Thinking Mode, we suppress chain-of-thought generation \upcite{b16}. Let $\mathcal{V}$ denote the vocabulary. We construct a binary logit mask $M \in \{-\infty, 0\}^{|\mathcal{V}|}$, enforcing $M_v = 0$ if $v \in \{t_{pass}, t_{fail}\}$ and $-\infty$ otherwise \upcite{b12}. The step $t=1$ distribution is:
\begin{equation}
P(y_1 \mid x) = \text{softmax}(z_1 \odot M) \label{eq:softmax}
\end{equation}
Enforcing $L_{max} = 1$ reduces decoding complexity to strictly $\mathcal{O}(|x|)$. This bounds routing overhead to the KV-cache prefill phase, accelerated by FlashAttention \upcite{b17}.

\section{Experimental Setup}

\subsection{Datasets and Adversarial Noise Injection Protocol}
We construct a 5,000-query corpus from Natural Questions and HotpotQA. To evaluate robustness, we apply an adversarial noise protocol ($\rho = 0.45$):
\begin{itemize}
\item \textbf{Hard Negatives:} Documents exhibiting high cosine similarity (rank 10-20 via BGE-M3) but lacking factual grounding \upcite{b18}.
\item \textbf{Conflicting Distractors:} Synthetic contexts containing explicitly falsified entities (e.g., altered dates or locations).
\end{itemize}

\subsection{Baselines and Model Configuration}
We evaluate three pipelines:
\begin{itemize}
\item \textbf{Naive RAG:} BGE-M3 piped directly to Llama-3-8B-Instruct ($T=0.1, \text{top-p}=0.9$).
\item \textbf{Heavy-CRAG:} Utilizes the \texttt{gpt-4o-mini} API as the evaluator. Fallback executes Tavily Search.
\item \textbf{Tiny-Critic RAG (Ours):} Deploys locally hosted \texttt{Qwen3-1.7B} \upcite{b11}. LoRA ($r=16, \alpha=32$) is applied to $W_{q,k,v,o}$. Trained for 15 epochs on an RTX 4090 using AdamW ($\eta = 3 \times 10^{-4}$).
\end{itemize}

\subsection{Evaluation Metrics}
Systemic performance is quantified via \upcite{b19}:
\begin{itemize}
\item \textbf{RAGAS Faithfulness ($F \in [0, 1]$):} Measures hallucination density relative to the context.
\item \textbf{Routing F1-Score:} Harmonic mean of precision and recall for identifying adversarial inputs.
\item \textbf{TTFT (ms):} System latency capturing routing and initial generation phases.
\item \textbf{CPQ (\$):} Cost Per 10k Queries. SLM inference utilizes AWS \texttt{g5.xlarge} pricing; GPT-4o-mini utilizes API rates.
\end{itemize}

\section{Experimental Results and Analyses}

\begin{table}[htbp]
\caption{System Performance and Cost Evaluation.}
\begin{center}
\setlength{\tabcolsep}{4pt} 
\small 
\begin{tabular}{|l|c|c|c|c|}
\hline
\textbf{Model} & \textbf{R-F1} & \textbf{Faithful.} & \textbf{TTFT (ms)} & \textbf{CPQ (\$)} \\
\hline
Naive RAG & N/A & 0.44 & 450 & 0.00 \\
Heavy-CRAG & \textbf{0.934} & \textbf{0.88} & 1235 & 3.00 \\
Tiny-Critic (Ours) & 0.912 & 0.86 & \textbf{492} & \textbf{0.06} \\
\hline
\multicolumn{5}{l}{$^{\mathrm{a}}$R-F1: Routing F1; Faithful.: RAGAS Faithfulness.}\\
\multicolumn{5}{l}{$^{\mathrm{b}}$CPQ: Explicit routing Cost Per 10k Queries in USD.}\\
\multicolumn{5}{l}{$^{\mathrm{c}}$CPQ estimations assume an average context of 2K tokens }\\
\multicolumn{5}{l}{under optimal batch utilization.}
\end{tabular}
\label{tab:metrics}
\end{center}
\end{table}

\subsection{Routing Efficacy and Noise Robustness}
The Tiny-Critic module demonstrates exceptional discriminative capability. The LoRA-tuned SLM achieves a Routing F1-Score of 0.912, statistically comparable to the 0.934 achieved by the heavyweight \texttt{gpt-4o-mini} baseline (Table \ref{tab:metrics}). 

\begin{figure}[htbp]
\centerline{\includegraphics[width=\columnwidth]{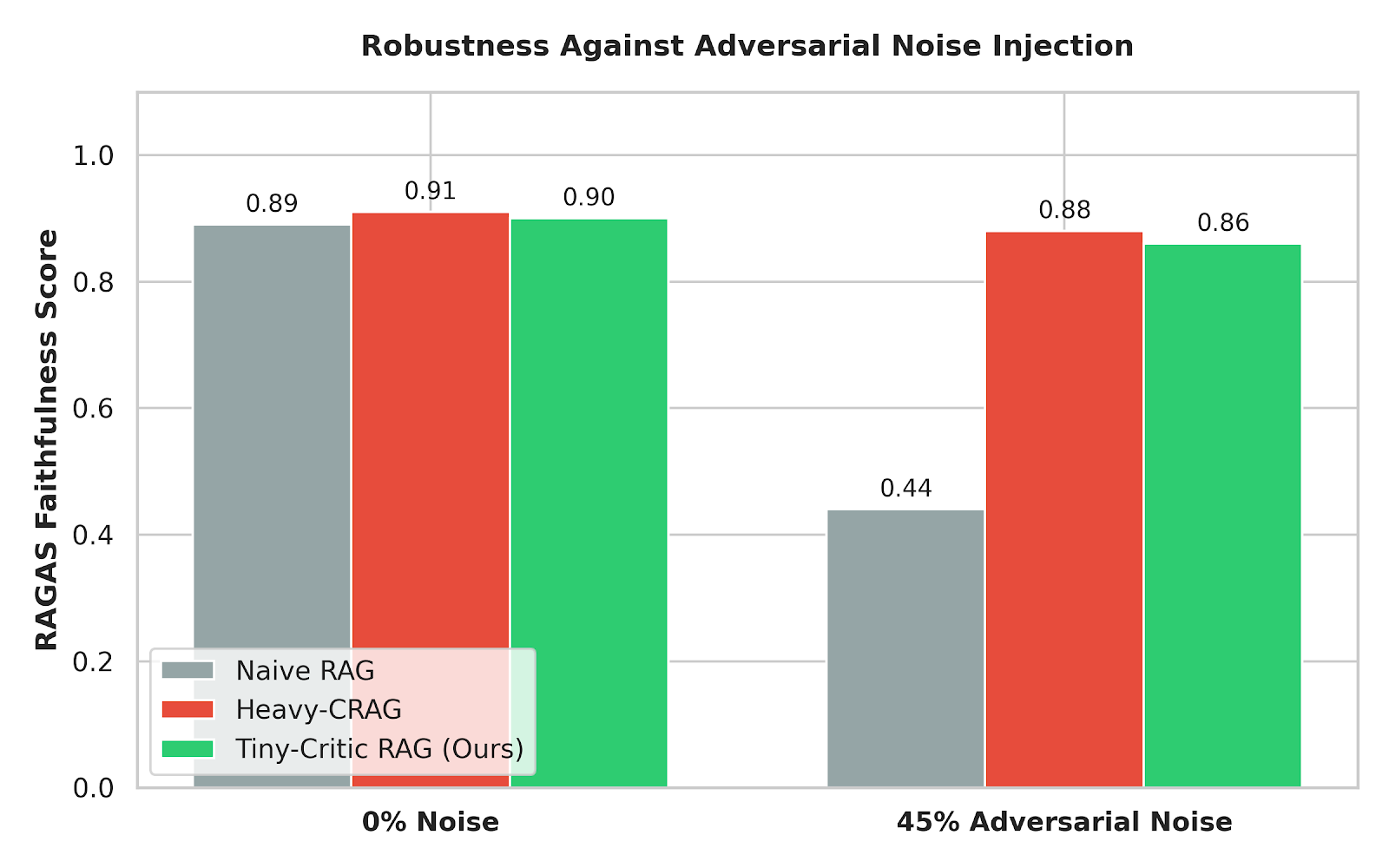}}
\caption{Robustness under $45\%$ adversarial noise. Tiny-Critic intercepts distractors, preventing catastrophic degradation.}
\label{fig:robustness}
\end{figure}

Naive RAG exhibits severe vulnerability. As illustrated in Figure \ref{fig:robustness}, under $\rho = 0.45$ noise, Faithfulness degrades from $0.89$ to $0.44$. By intercepting hard negatives, Tiny-Critic RAG sustains a Faithfulness score of $0.86$, completely bypassing contamination cascades.

\subsection{Latency and Cost Profiling}
Tiny-Critic RAG profoundly mitigates both explicit evaluation overhead and implicit agentic costs.

\begin{figure}[htbp]
\centerline{\includegraphics[width=\columnwidth]{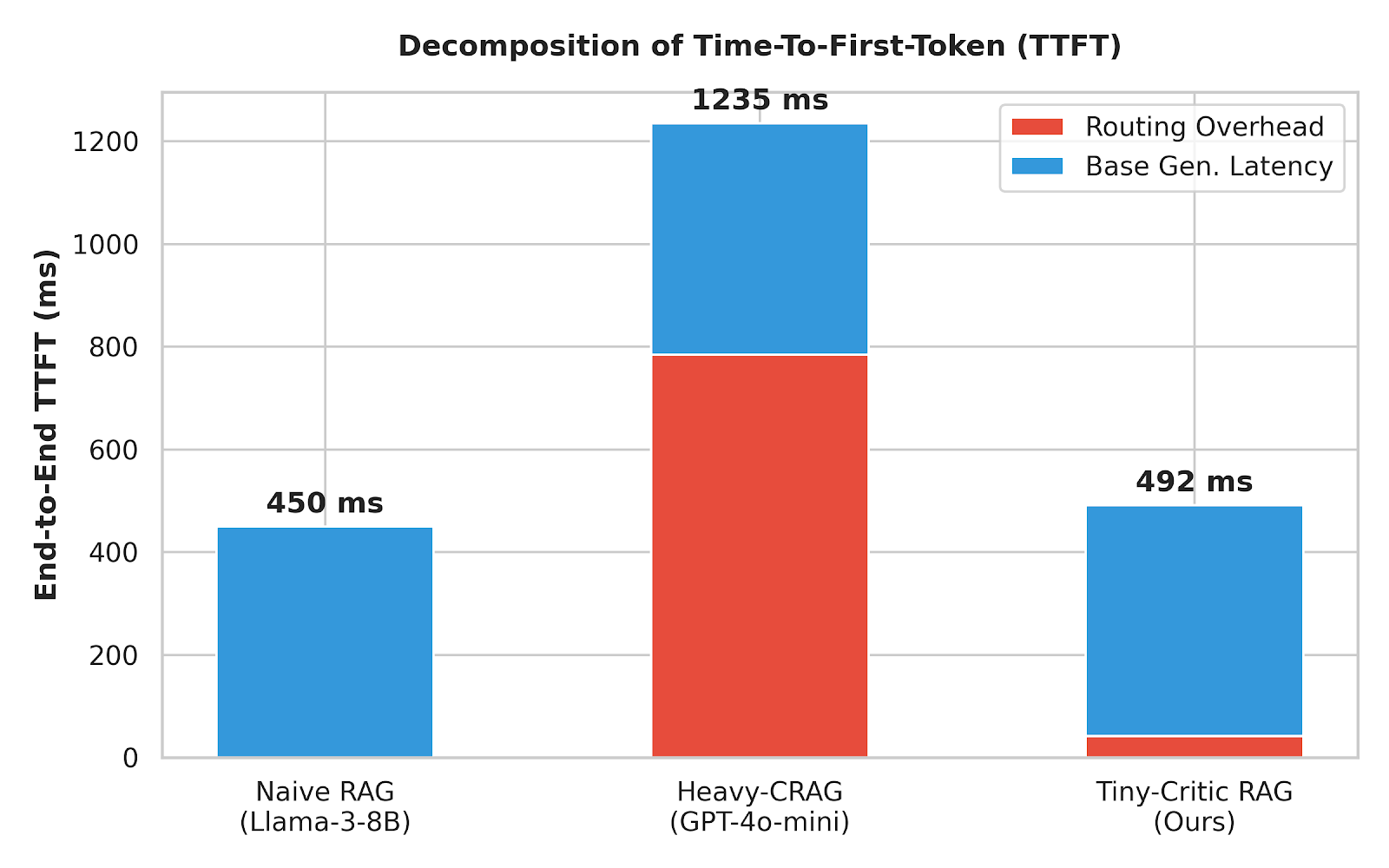}}
\caption{Routing TTFT comparison. Tiny-Critic RAG achieves a 94.6\% reduction in routing overhead compared to Heavy-CRAG.}
\label{fig:latency}
\end{figure}

\begin{itemize}
\item \textbf{Latency:} As shown in Figure \ref{fig:latency}, Heavy-CRAG introduces a 785 ms routing overhead. Through constrained decoding and Non-Thinking inference, the local Tiny-Critic processes routing in merely 42 ms. This represents a 94.6\% reduction in routing overhead, making reflection nearly imperceptible.
\item \textbf{Financial Cost:} Processing 10,000 queries incurs an explicit CPQ of \$3.00 for Heavy-CRAG. Conversely, Tiny-Critic incurs merely \$0.06. Furthermore, preventing multi-hop reasoning over faulty evidence saves an estimated \$1.20 per 10k queries in implicit token waste compared to standard ReAct workflows.
\end{itemize}

\begin{figure}[htbp]
\centerline{\includegraphics[width=\columnwidth]{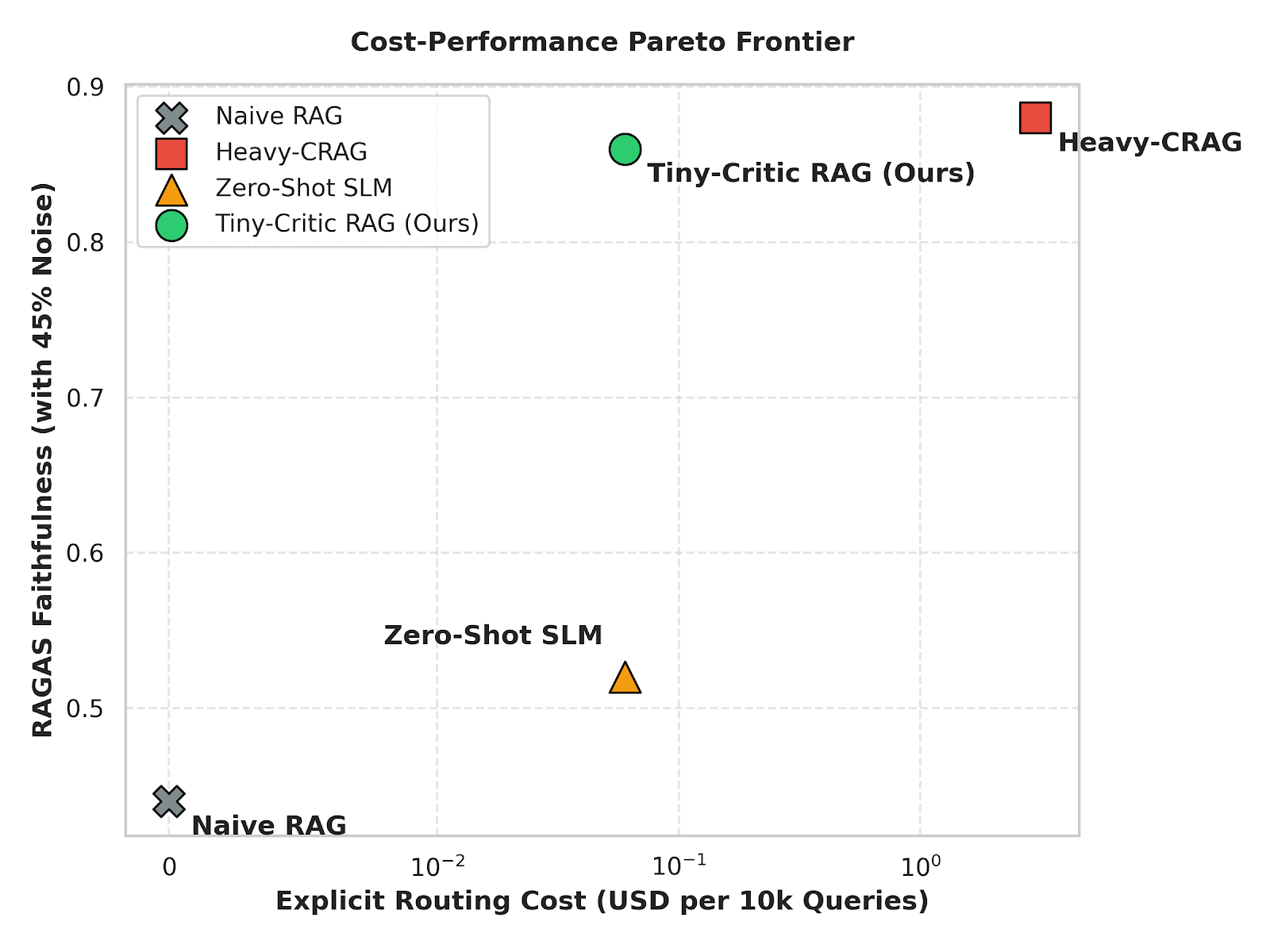}}
\caption{Cost-Performance Pareto Frontier. Tiny-Critic optimally balances noise robustness with near-zero marginal evaluation cost.}
\label{fig:pareto}
\end{figure}

The Pareto distribution (Figure \ref{fig:pareto}) confirms our approach matches SOTA Agentic RAG robustness at near-zero marginal evaluation costs.

\subsection{Ablation on Adaptation Strategies}
A zero-shot Qwen3-1.7B fails to adhere to binary boundaries, yielding a False Positive Rate (FPR) of 38.2\% due to inherent sycophancy and verbosity. LoRA training explicitly bounds the latent space, overriding sycophancy and reducing FPR to 4.1\%, confirming task-specific alignment is required for deterministic routing.

\section{Conclusion}
This research introduced Tiny-Critic RAG, a highly optimized architectural framework engineered to decisively resolve the critical latency and operational cost bottlenecks inherent in reflective Agentic RAG systems. We comprehensively demonstrated that in autonomous agent scenarios, inaccurate retrieval evidence does not merely induce localized factual hallucinations; rather, it fundamentally degrades system economics by triggering implicit multi-hop reasoning spirals and highly redundant tool calls. By systematically decoupling the evaluation mechanism from heavy, cloud-based LLMs and integrating a parameter-efficient, LoRA-adapted SLM combined with constrained decoding and non-thinking execution modes, we successfully established an ultra-low latency routing module. 

Empirical analyses rigorously validate that Tiny-Critic RAG achieves a staggering 94.6\% reduction in evaluation TTFT and a 98\% reduction in explicit operational CPQ, all while maintaining an exceptional Routing F1-Score of 0.912 against complex adversarial noise. Future work will investigate extending this lightweight routing mechanism via standardized Model Context Protocols (MCP) to govern multi-modal evidence retrieval utilizing highly quantized vision-language models.


\end{document}